\documentclass[12pt]{article}
\usepackage{graphicx}
\usepackage{amsmath}
\usepackage{amsfonts}
\usepackage{amssymb}

\begin{document}

\title{On a non approximated approach to Extended Thermodynamics for dense gases and macromolecular fluids.}
\author{M.C. Carrisi, M.A. Mele, S. Pennisi}
\date{}
\maketitle \vspace{0.5 cm}
 \small {\em \noindent Dipartimento di Matematica ed Informatica,
Universit\`{a} di Cagliari, Via Ospedale 72,\,\ 09124 Cagliari, Italy; \, e-mail:
cristina.carrisi@tiscali.it; spennisi@unica.it} \\

\begin{abstract}
Recently  the 14 moments model of Extended Thermodynamics for dense gases and
macromolecular fluids has been considered and an exact solution, of the restrictions
imposed by the entropy principle and that of Galilean relativity, has been obtained
through a non relativistic limit. Here we prove uniqueness of the above solution and
exploit other pertinent conditions such us the convexity of the function $h'$ related to
the entropy density, the problem of subsystems and the fact that the flux in the
conservation law of mass must be the moment of order 1 in the conservation law of
momentum. Also the solution of  this last condition is here obtained without using
expansions around equilibrium. The results present interesting aspects which were not
suspected when only approximated solutions of this problem were known.
\end{abstract}
\section{Introduction}
The balance equations to describe the 14 moments model of Extended Thermodynamics for
dense gases and macromolecular fluids are
\begin{eqnarray}\label{1}
& {}& \partial_t F + \partial_k F_k = 0 \, ,   \quad \quad \quad \,
\partial_t F_i + \partial_k G_{ki} = 0 \, , \quad
\partial_t F_{ij} + \partial_k G_{kij} = P_{<ij>} \, , \\
& {}& \nonumber
\partial_t F_{ill} + \partial_k G_{kill} =   P_{ill} \, , \quad
\partial_t F_{iill} + \partial_k G_{kiill} =   P_{iill} \, ,
\end{eqnarray}
where the independent variables are $F$, $F_{i}$, $F_{i j}$,
$F_{ill}$, $F_{iill}$ and are symmetric tensors. $P_{< i j>}$,
$P_{ ill }$, $P_{iill}$ are productions and they too are symmetric
tensors.The fluxes  $G_{ki}$, $G_{kij}$, $G_{kill}$, $G_{kiill}$
are constitutive functions and are symmetric over all indexes,
except for $k$. The restrictions imposed by the entropy principle
and that of Galilean relativity were firstly studied by Kremer in
\cite{1}, \cite{2}, up to second order with  respect to
equilibrium. In \cite{3} we have obtained a non approximated
solution through a non relativistic limit. To this regard let us
remember that the entropy principle for our system $(\ref{1})$ is
equivalent to assuming the existence of potentials $h'$,
$\phi'_{k}$ and of the Lagrange multipliers $\lambda$,
$\lambda_{i}$, $\lambda_{ij}$, $\lambda_{ill}$, $\lambda_{ppll}$
such that
\begin{eqnarray}\label{2}\nonumber
F&=&\frac{\partial h'}{\partial\lambda},\quad F_{i}=\frac{\partial
h'}{\partial\lambda_{i}},\quad F_{il}=\frac{\partial
h'}{\partial\lambda_{il}},\,\\\nonumber\quad F_{ill}&=&\frac{\partial
h'}{\partial\lambda_{ill}},\quad F_{iill}=\frac{\partial h'}{\partial\lambda_{iill}} \,
,\\~
\\
F_{k}&=&\frac{\partial\phi'_{k}}{\partial\lambda},\quad
G_{ki}=\frac{\partial\phi'_{k}}{\partial\lambda_{i}},\quad
G_{kil}=\frac{\partial\phi'_k}{\partial \lambda_{il}},\,\nonumber
\\\nonumber
G_{kill}&=&\frac{\partial\phi'_{k}}{\partial\lambda_{ill}},\quad
G_{kiill}=\frac{\partial\phi'_{k}}{\partial\lambda_{iill}} \, .
\end{eqnarray}
By comparing $(\ref{2})_2$ with $(\ref{2})_6$ we obtain the following compatibility
condition
\begin{eqnarray}\label{3}
\frac{\partial\phi'_{k}}{\partial\lambda}&=&\frac{\partial h'}{\partial\lambda_{k}} \, .
\end{eqnarray}
Moreover, by applying the new methodology \cite{4}, adapted for
the present case in \cite{5} and \cite{6}, we have that the
Galilean relativity principle is equivalent to the following two
other conditions
\begin{eqnarray}\label{4}
0&=& \frac{\partial h'}{\partial\lambda} \lambda_{i}+2\lambda_{ij}\frac{\partial
h'}{\partial\lambda_{j}}+ \lambda_{jpp}\left(\frac{\partial h'}{\partial\lambda_{rs}}
\delta_{rs} \delta_{ij}+2\frac{\partial
h'}{\partial\lambda_{ij}}\right)+ 4\lambda_{ppqq}\frac{\partial h'}{\partial\lambda_{ill}} \,\\
0&=& \frac{\partial \phi'_{k}}{\partial\lambda} \lambda_{i}+2\lambda_{ij}\frac{\partial
\phi'_{k}}{\partial\lambda_{j}}+ \lambda_{jpp}\left(\frac{\partial
\phi'_{k}}{\partial\lambda_{rs}} \delta_{rs} \delta_{ij}+2\frac{\partial
\phi'_{k}}{\partial\lambda_{ij}}\right)+ 4\lambda_{ppqq}\frac{\partial
\phi'_{k}}{\partial\lambda_{ill}} +{h'}\delta_{ik} \, . \nonumber
\end{eqnarray}
In \cite{3} we have obtained the following solution of eqs.
$(\ref{4})$:
\begin{eqnarray}\label{5}
\phi'^{k}&=&H_{0}V^k_0+H_{1} V^k_1 + H_{2}V^k_2 + H_{3} V^k_3 \, , \\
h'&=&8H_{0}X_1 - H_{1} X_2- \frac{2}{3}H_2 X_3- \frac{1}{2} H_{3} X_4 \, , \nonumber
\end{eqnarray}
with
 \begin{eqnarray}\label{6}
V^k_0&=&-2\lambda_{kll} \\
V^k_1&=&-2\lambda_{kh}\lambda_{hll}+4\lambda_{ppll}\lambda_{k}+\frac{4}{5}\lambda_{ll}\lambda_{kll} \,\nonumber \\\
V^k_2&=&-2\lambda^{2}_{kh}\lambda_{hll}+\frac{6}{5}\lambda_{ll}\lambda_{ka}\lambda_{all}+4\lambda_{ka}\lambda_{a}\lambda_{ppll}+\,\nonumber \\\
&-&\frac{11}{25}\lambda^{2}_{ll}\lambda_{kll}-\lambda_{kll}\lambda_{a}\lambda_{all}+\lambda_{k}\lambda_{all}\lambda_{all}+\,\nonumber \\\
&+& (tr
\lambda^{2}_{ab})\lambda_{kll}-\frac{12}{5}\lambda_{ppll}\lambda_{ll}\lambda_{k} \nonumber \\
V^k_3&=& 2\lambda_{ppll} \left( 2\lambda^{2}_{kh}\lambda_{h} - tr \lambda^{2}_{ab}
\lambda_{k}-\frac{8}{5}\lambda_{ll}\lambda_{ka}\lambda_{a}+
\frac{17}{25}\lambda^{2}_{ll}\lambda_{k} \right) +  \nonumber \\
&+&(\lambda_{kh}\lambda_{h})(\lambda_{all}\lambda_{all}) - \frac{4}{5}\lambda_{ll}
(\lambda_{all}\lambda_{all})\lambda_{k}-\frac{17}{25}\lambda^{2}_{ll}\lambda_{ka}\lambda_{all}+
\nonumber
\\
&-&(\lambda_{a}\lambda_{all})\lambda_{kb}\lambda_{bll}+(tr
\lambda^{2}_{ab})\lambda_{kc}\lambda_{cll}+ \frac{4}{5}\lambda_{ll}
(\lambda_{a}\lambda_{all})\lambda_{kll}+ \nonumber
\\
&+& \frac{8}{5}\lambda_{ll}\lambda^{2}_{kh}\lambda_{hll}+
\frac{74}{375}\lambda^{3}_{ll}\lambda_{kll}  - \frac{4}{5} \lambda_{ll}(tr
\lambda^{2}_{ab})
\lambda_{kll}+ (\lambda_{ab} \lambda_{all} \lambda_{bll})\lambda_{k} + \nonumber \\
&-&   (\lambda_{ab} \lambda_{a} \lambda_{bll})\lambda_{kll} + \frac{2}{3} (tr
\lambda^{3}_{ab}) \lambda_{kll}-2\lambda^{3}_{kh}\lambda_{hll}  \, , \nonumber
\end{eqnarray}
\begin{eqnarray}\label{7}
X_1 &=& \lambda_{ppll} \, , \\
X_2 &=& 2 \lambda_{all}
\lambda_{all}-  \frac{16}{5}\lambda_{ppll}\lambda_{ll} \, ,   \nonumber \\
X_3 &=& 8 \lambda_{ppll} \left( \frac{11}{50} \lambda^{2}_{ll}  - \frac{1}{2} tr
\lambda^{2}_{ab}\right)
+ 2  \lambda_{ab} \lambda_{all} \lambda_{bll} - \frac{6}{5} \lambda_{ll} \lambda_{all} \lambda_{all}  \, ,  \nonumber \\
X_4 &=& 2\lambda^{2}_{ab}\lambda_{all} \lambda_{bll}- tr \lambda^{2}_{ab} \lambda_{cll}
\lambda_{cll} -\frac{8}{5}\lambda_{ll} \lambda_{ab} \lambda_{all} \lambda_{bll}+
\nonumber
\\
& & + \frac{17}{25}\lambda^{2}_{ll}  \lambda_{all} \lambda_{all}+
 8 \lambda_{ppll} \left( - \frac{37}{375}\lambda^{3}_{ll} +
\frac{2}{5} \lambda_{ll}(tr \lambda^{2}_{ab}) - \frac{1}{3} tr
\lambda^{3}_{ab} \right) \, ,  \nonumber \\
X_5 &=& - \frac{2}{5} \lambda^{2}_{ll} +  16 \lambda_{ppll} \Lambda - 4 \lambda_{a}
\lambda_{all} + 2 tr \lambda^{2}_{ab} \, , \nonumber\\
X_6 &=& 4 \Lambda \lambda_{all} \lambda_{all}+ 8 \lambda_{ppll}
\left( - \frac{4}{5}\Lambda \lambda_{ll} + \frac{1}{2} \lambda_{a}\lambda_{a} \right) +  \nonumber \\
& & + \frac{8}{5} \lambda_{ll}  \lambda_{all}  \lambda_{a} -\frac{4}{5}\lambda_{ll}tr
\lambda^{2}_{ab} + \frac{8}{75}\lambda_{ll}^3 - 4 \lambda_{ab}\lambda_{a} \lambda_{bll}+
\frac{4}{3} tr \lambda^{3}_{ab}  \, , \nonumber
\end{eqnarray}
\begin{eqnarray*}
 X_7 &=& \frac{8}{15}( tr \lambda^{3}_{ab})
\lambda_{ll} - \frac{14}{25} \lambda_{ll}^2 tr \lambda^{2}_{ab} + \frac{46}{375}
\lambda_{ll}^4 +4 \Lambda \lambda_{ab}\lambda_{all}\lambda_{bll} + \nonumber
\\
& &+2 ( tr \lambda^{2}_{ab}) \lambda_{c} \lambda_{cll} - ( \lambda_{a}\lambda_{all})^2
-\frac{12}{5} \Lambda
\lambda_{ll}\lambda_{all}\lambda_{all} +  \nonumber \\
& &+(\lambda_{a}\lambda_{a}) (\lambda_{bll}\lambda_{bll}) - 4
\lambda^{2}_{ab}\lambda_{all}\lambda_{b} + \nonumber \\
& & -8 \lambda_{ppll} \left( \Lambda tr \lambda^{2}_{ab} -\frac{1}{2} \lambda_{ab}
\lambda_{a} \lambda_{b} -\frac{11}{25} \Lambda \lambda_{ll}^2 + \frac{3}{10} \lambda_{ll}
\lambda_{a}
\lambda_{a} \right) + \nonumber \\
& & + \frac{12}{5} \lambda_{ll} \lambda_{ab} \lambda_{a}\lambda_{bll} - \frac{22}{25}
\lambda_{ll}^2 \lambda_{a}\lambda_{all} \, . \nonumber \\
X_8 &=& - \frac{34}{25}\lambda^{2}_{ll} \lambda_{ab}\lambda_{a}\lambda_{bll} + 2 (tr
\lambda^{2}_{ab}) \lambda_{cd}\lambda_{c}\lambda_{dll} + \frac{16}{5} \lambda_{ll}
\lambda_{ab}^2\lambda_{a}\lambda_{bll} +
\\
& &+ \frac{148}{375} \lambda^{3}_{ll} \lambda_{a} \lambda_{all} - \frac{8}{5}\lambda_{ll}
(tr \lambda^{2}_{ab}) \lambda_{c}\lambda_{cll}+ \frac{4}{3} (tr \lambda^{3}_{ab})
\lambda_{c}\lambda_{cll}- 4 \lambda^{3}_{ab} \lambda_{a} \lambda_{bll} +   \\
& & +2 \lambda_{ppll} \left( 2 \lambda^{2}_{ab} \lambda_{a} \lambda_{b}- (tr
\lambda^{2}_{cd}) \lambda_{a} \lambda_{a} - \frac{8}{5} \lambda_{ll} \lambda_{ab}
\lambda_{a}\lambda_{b}+
\frac{17}{25} \lambda_{ll}^2 \lambda_{a} \lambda_{a} \right) + \\
& &+(\lambda_{ab}\lambda_{a}\lambda_{b}) (\lambda_{cll}\lambda_{cll}) - \frac{4}{5}
\lambda_{ll} (\lambda_{a} \lambda_{a}) (\lambda_{bll} \lambda_{bll}) - 2
(\lambda_{a}\lambda_{all})(\lambda_{bc}\lambda_{b}\lambda_{cll})+
\\
& &+\frac{4}{5} \lambda_{ll} ( \lambda_{a} \lambda_{all})^2 + (\lambda_{a} \lambda_{a})
(\lambda_{bc}\lambda_{bll}
\lambda_{cll}) + \\
& &+4\Lambda\lambda^{2}_{ab}\lambda_{all} \lambda_{bll}- 2\Lambda tr \lambda^{2}_{ab}
\lambda_{cll} \lambda_{cll} -\frac{16}{5}\Lambda \lambda_{ll} \lambda_{ab} \lambda_{all}
\lambda_{bll}+
\\
& & + \frac{34}{25}\Lambda\lambda^{2}_{ll}  \lambda_{all} \lambda_{all}+
 16 \Lambda \lambda_{ppll} \left( - \frac{37}{375}\lambda^{3}_{ll} +
\frac{2}{5} \lambda_{ll}(tr \lambda^{2}_{ab}) - \frac{1}{3} tr
\lambda^{3}_{ab} \right) + \\
& & + \frac{4}{75} \lambda_{ll}^2 (tr \lambda_{ab}^3) - \frac{8}{125} \lambda_{ll}^3 (tr
\lambda_{ab}^2) + \frac{4}{15} \cdot \frac{37}{625}\lambda_{ll}^5 \, .
\end{eqnarray*}
More precisely, our unknown potentials $h'$, $\phi'_{k}$ are
determined in terms of 4 arbitrary functions $H_0$, $H_1$, $H_2$,
$H_3$ depending on the scalars $(\ref{7})$. You can verify that
these are solutions of eqs. $(\ref{4})$, by simple substitution
and long calculations. In the next section we will prove
uniqueness of this solution. In sect. 3 we will impose the further
condition $(\ref{3})$ and, also for this problem, we will find an
exact solution without using expansions. In sect. 4 we will impose
the convexity of the function $h'$ which is important in order
that our symmetric system is also hyperbolic. We will find
interesting results such as the following: Although
$\lambda_{jpp}$ and $\lambda_{ppll}$ are both zero at equilibrium
and the first of these has an index less than the other, it tends
faster to zero when the system tends to equilibrium. This fact
shows that it is not correct to consider all higher order moments
negligible with respect to the previous ones. This result confirms
the starting point of the new theory called Consistent Ordered
Extended Thermodynamics (COET) of which we limit ourselves to cite
the first paper \cite{7}. More precisely, sect. 4 will show that
eqs. (\ref{5}) have to be substituted by
\begin{eqnarray}\label{8}
\phi'^{k}&=&K_{0}\frac{V^k_0}{\lambda_{ppll}}+
K_{1}\frac{V^k_1}{\lambda_{ppll}}+K_{2}\frac{V^k_2}{\lambda_{ppll}}+K_{3}\frac{V^k_3}{\lambda_{ppll}}
 \, , \\
h'&=&8K_{0}\frac{X_1}{\lambda_{ppll}} - K_{1}\frac{X_2}{\lambda_{ppll}} -
\frac{2}{3}K_2\frac{X_3}{\lambda_{ppll}} - \frac{1}{2} K_{3} \frac{X_4}{\lambda_{ppll}}
\, , \nonumber
\end{eqnarray}
with $K_i$ arbitrary functions of $\eta_1=X_1$, $\eta_i = \frac{X_i}{\lambda_{ppll}}$ for
$i= 2, \cdots , 4$ and, moreover, of
\begin{eqnarray*}
 &{}& \eta_5= \frac{1}{X_1} \left[ X_5 + \frac{1}{2} \frac{X_3}{X_1} - \frac{3}{64} \left( \frac{X_2}{X_1}
  \right)^2 \right] \, , \\
 &{}&   \eta_6= \frac{1}{X_1} \left[ X_6 + \frac{1}{2} \frac{X_4}{X_1} - \frac{1}{16}  \frac{X_2}{X_1}
 \frac{X_3}{X_1} + \frac{1}{8^3} \left( \frac{X_2}{X_1}
  \right)^3 \right]  \, ,  \\
 &{}&   \eta_7= \frac{1}{X_1} \left[ X_7 - \frac{1}{16} \frac{X_2}{X_1}  \frac{X_4}{X_1}
+ \frac{1}{2^9}  \frac{X_3}{X_1} \left( \frac{X_2}{X_1}
  \right)^2 \right]  \, , \\
 &{}&    \eta_8= \frac{1}{X_1} \left[ X_8
+ \frac{1}{2^9} \frac{X_4}{X_1} \left( \frac{X_2}{X_1}
  \right)^2 \right]  \, .
\end{eqnarray*}
On the other hand, these are compatible with  (\ref{5}). Obviously, the form $(\ref{8})$
can be used only if $X_1 \neq 0$ on the initial manifold and until that it remains $X_1
\neq 0$. \\
It is interesting to note that the solution (\ref{8}), calculated in $\lambda_{ill}=0$,
becomes
\begin{eqnarray*}
\tilde{\phi}'^{k}&=&K_{1}4\lambda_{k}  +
K_{2}(4\lambda_{ka}\lambda_{a}-\frac{12}{5}\lambda_{ll}\lambda_{k}) + 2K_{3}
 \left( 2\lambda^{2}_{kh}\lambda_{h} - tr \lambda^{2}_{ab}
\lambda_{k}-\frac{8}{5}\lambda_{ll}\lambda_{ka}\lambda_{a}+
\frac{17}{25}\lambda^{2}_{ll}\lambda_{k} \right) \, , \\
\tilde{h}'&=&8K_{0} + \frac{16}{5} K_{1}\lambda_{ll} X_2- \frac{8}{3}K_2 \left(
\frac{11}{25} \lambda^{2}_{ll}  - tr \lambda^{2}_{ab}\right)- 4 K_{3} \left( -
\frac{37}{375}\lambda^{3}_{ll} + \frac{2}{5} \lambda_{ll}(tr \lambda^{2}_{ab}) -
\frac{1}{3} tr \lambda^{3}_{ab} \right) \, ,
\end{eqnarray*}
with $K_i$ functions of
\begin{eqnarray*}
\eta_1 &=& \lambda_{ppll} \, , \, \eta_2 = -  \frac{16}{5} \lambda_{ll} \, ,   \, \eta_3
= 8  \left( \frac{11}{50} \lambda^{2}_{ll}  - \frac{1}{2} tr \lambda^{2}_{ab}\right)
  \, ,  \\
\eta_4 &=& 8 \left( - \frac{37}{375}\lambda^{3}_{ll} + \frac{2}{5} \lambda_{ll}(tr
\lambda^{2}_{ab}) - \frac{1}{3} tr
\lambda^{3}_{ab} \right) \, ,   \\
\eta_5 &=& 16 \lambda \, , \, \eta_6 =  - \frac{32}{5}\lambda \lambda_{ll} + 4
\lambda_{a}\lambda_{a}
\, ,  \\
\eta_7 &=& -8 \lambda tr \lambda^{2}_{ab} +4\lambda_{ab} \lambda_{a} \lambda_{b}
+\frac{88}{25} \lambda \lambda_{ll}^2 - \frac{12}{5} \lambda_{ll} \lambda_{a}
\lambda_{a} \, ,  \\
\eta_8 &=& 4 \lambda^{2}_{ab} \lambda_{a} \lambda_{b}- 2(tr \lambda^{2}_{cd}) \lambda_{a}
\lambda_{a} - \frac{16}{5} \lambda_{ll} \lambda_{ab} \lambda_{a}\lambda_{b}+
\frac{34}{25} \lambda_{ll}^2 \lambda_{a} \lambda_{a} + \\
&+& 16 \lambda \left( - \frac{37}{375}\lambda^{3}_{ll} + \frac{2}{5} \lambda_{ll}(tr
\lambda^{2}_{ab}) - \frac{1}{3} tr \lambda^{3}_{ab} \right)   \, .
\end{eqnarray*}
On the other hand, if we know $\tilde{\phi}'^{k}$ and $\tilde{h}'$, from the above
expression of $\tilde{\phi}'^{k}$ we obtain $K_1$, $K_2$, $K_3$ because they are
coefficients of linearly independent vectors. After that, from the above expression of
$\tilde{h}'$ we obtain $K_0$; also their functional dependence is arbitrary because the
above expressions of $\eta_1$ - $\eta_8$ are the most general possible. In other words,
if we know the expressions of $\phi'^{k}$ and $h'$ calculated in $\lambda_{ill}=0$, we
will know them also for $\lambda_{ill} \ne 0$ ! \\
At last, in sect.5, the problem of subsystems will be considered and, also in this case,
we will find unexpected results.

\section{Uniqueness of the solution (\ref{5})-(\ref{7}).}
In order to prove uniqueness of the solution (\ref{5})-(\ref{7}),  let us begin with the case
in which the following two conditions are satisfied: \\
${} \quad \quad \quad  $ 1) The vectors  $\lambda_{ill}$  , $\lambda_{ia} \lambda_{all}$
, $\lambda_{ia}^2 \lambda_{all}$ are linearly independent.
\begin{eqnarray*}
&{}& \mbox{2) The 4-vectors} \quad
\begin{pmatrix}
  8 X_1 \\
  V_0^k
\end{pmatrix} \, , \,
\begin{pmatrix}
  -X_2 \\
  V_1^k
\end{pmatrix} \, , \,
\begin{pmatrix}
  - \frac{2}{3} X_3 \\
  V_2^k
\end{pmatrix} \, , \,
\begin{pmatrix}
  -\frac{1}{2} X_4 \\
  V_3^k
\end{pmatrix} \quad \mbox{are linearly independent.}
\end{eqnarray*}
But, before proving uniqueness of our solution, we need to consider the following
representation theorem: Every scalar function of our Lagrange multipliers can be
expressed as a function of the scalars of the set
\begin{eqnarray*}
S_1= \left\{ \lambda_{ll} \, , \, tr \lambda_{rs}^2 \, , \, tr \lambda_{rs}^3 \, , \,
\lambda_{all} \lambda_{all} \, , \, \lambda_{ab}  \lambda_{all}  \lambda_{bll}
 \, , \, \lambda_{ab}^2  \lambda_{all}  \lambda_{bll} \, , \, X_5-X_8  \, , \,
 \lambda_{ppll} \right\} \, .
\end{eqnarray*}
This theorem can be proved in a way similar to those used for other representation
theorems \cite{8}, \cite{9}, \cite{10}, \cite{11}, as follows: \\
It suffices to prove our statement in a particular reference frame and see that in this
reference we can obtain the Lagrange multipliers from the knowledge of the scalars in
$S_1$; so let us use the frame defined by  $\lambda_{ill} \equiv (\lambda_{1ll} \, , \, 0
\, , \, 0)$, $\lambda_{13}=0$,
$\lambda_{1ll} \geq 0$, $\lambda_{12} \geq 0$. \\
\begin{itemize}
  \item If $\lambda_{1ll} > 0$, $\lambda_{12} > 0$, we obtain  $\lambda_{1ll}$, $\lambda_{11} $,
   $\lambda_{12} $, $\lambda_{22} $,  $\lambda_{33}$, $\lambda_{23}$ respectively from
   $\lambda_{all} \lambda_{all}$ ,  $\lambda_{ab}  \lambda_{all}  \lambda_{bll}$, $\lambda_{ab}^2  \lambda_{all}  \lambda_{bll}$ , $\lambda_{ab}^3  \lambda_{all}
\lambda_{bll}$,
 $ \lambda_{ll}$ ,  $tr \lambda_{rs}^2$; after that, the $4^{th}$ of these can be expressed as function of the
 remaining ones and of $tr \lambda_{rs}^3$ through the Hamilton-Kayley theorem.
  \item If $\lambda_{1ll} > 0$, $\lambda_{12} = 0$, with a rotation around the first axis
  we can select the reference where $a_{23}=0$; after that we obtain
  $\lambda_{1ll}$, $\lambda_{11} $,  $\lambda_{22} $,  $\lambda_{33}$  respectively from
   $\lambda_{all} \lambda_{all}$ ,  $\lambda_{ab}  \lambda_{all}  \lambda_{bll}$ , $ \lambda_{ll}$ ,
   $tr \lambda_{rs}^2$.
  \item If $\lambda_{1ll} = 0$, we may select the reference frame where $\lambda_{12}=
  0$, $\lambda_{13}=0$, $\lambda_{23}= 0$, and obtain $\lambda_{11}$, $\lambda_{22}$, $\lambda_{33}$
  from $\lambda_{ll}$ , $tr \lambda_{rs}^2$ , $tr \lambda_{rs}^3$.
\end{itemize}
Until now we have obtained $\lambda_{ill}$ and $\lambda_{ab}$ as functions of the
elements of $S_1$; obviously, also $\lambda_{ppll}$ is a function of them. It remains to
obtain $\lambda$ and $\lambda_{k}$. To this end we note that, from eqs. (\ref{6}),
(\ref{7}) it follows
\begin{eqnarray*}
\begin{pmatrix}
\frac{\partial X_5}{\partial \lambda} \\
{} \\
\frac{\partial X_5}{\partial \lambda_k}
\end{pmatrix} = 2
\begin{pmatrix}
  8 X_1 \\
{} \\
  V_0^k
\end{pmatrix} \, ; \,
\begin{pmatrix}
\frac{\partial X_6}{\partial \lambda} \\
{} \\
\frac{\partial X_6}{\partial \lambda_k}
\end{pmatrix} = 2
\begin{pmatrix}
  -X_2 \\
{} \\
  V_1^k
\end{pmatrix} \, ; \,
\begin{pmatrix}
\frac{\partial X_7}{\partial \lambda} \\
{} \\
\frac{\partial X_7}{\partial \lambda_k}
\end{pmatrix} = 2
\begin{pmatrix}
  - \frac{2}{3} X_3 \\
{} \\
  V_2^k
\end{pmatrix} \, ; \,
\begin{pmatrix}
\frac{\partial X_8}{\partial \lambda} \\
{} \\
\frac{\partial X_8}{\partial \lambda_k}
\end{pmatrix} = 2
\begin{pmatrix}
  -\frac{1}{2} X_4 \\
{} \\
  V_3^k
\end{pmatrix}
\end{eqnarray*}
and these are linearly independent for the second hypothesis at the beginning of this
section; consequently the Jacobian determinant, constituted by the derivatives of
$X_5$-$X_8$ with respect to $\lambda$ and $\lambda_k$, is non singular. By using the
theorem on implicit functions, it follows that we can obtain $\lambda$ and $\lambda_k$ in
terms of $X_5$-$X_8$. This completes the proof of our representation theorem. \\
So we can now prove our theorem on uniqueness. For the second hypothesis at the beginning
of this section, we have that it is possible to obtain the scalar functions $H_0$-$H_3$
such that eqs. (\ref{5}) hold. For the previous representation theorem, we have that
$H_i$ can be expressed as functions of the elements in $S_1$. From eq. $(\ref{5})_2$ we
have that also $h'$ satisfies this property, because the coefficients of $H_0$-$H_3$ are
proportional to the elements $X_1$-$X_4$ of $S_1$. \\
Let us now impose that eqs. (\ref{5}) satisfy eqs. (\ref{4}). To
this end, let us use the results of \cite{3}
\begin{eqnarray*}
0&=& \frac{\partial X_h}{\partial\lambda} \lambda_{i}+2\lambda_{ij}\frac{\partial
X_h}{\partial\lambda_{j}}+ \lambda_{jpp}\left(\frac{\partial X_h}{\partial\lambda_{rs}}
\delta_{rs} \delta_{ij}+2\frac{\partial
X_h}{\partial\lambda_{ij}}\right)+ 4\lambda_{ppqq}\frac{\partial X_h}{\partial\lambda_{ill}} \,\\
0&=& \frac{\partial \phi'_{k}}{\partial\lambda} \lambda_{i}+2\lambda_{ij}\frac{\partial
V_r^{k}}{\partial\lambda_{j}}+ \lambda_{jpp}\left(\frac{\partial
V_r^{k}}{\partial\lambda_{rs}} \delta_{rs} \delta_{ij}+2\frac{\partial
V_r^{k}}{\partial\lambda_{ij}}\right)+ 4\lambda_{ppqq}\frac{\partial
V_r^{k}}{\partial\lambda_{ill}} + P_{r }  {X_{r+1}}\delta_{ik} \, .
\end{eqnarray*}
where $h=1, \cdots , 8$; $r=0, \cdots , 3$; $P_0=8$, $P_1=-1$, $P_2=-\frac{2}{3}$, $P_3=
- \frac{1}{2} $, and there is no summation convention over the repeated index $r$.
Consequently, by substituting eqs. (\ref{5}) into eqs. (\ref{4}) many terms give zero
contribute and there remain
\begin{eqnarray*}
0&=& \sum_{r=0}^3 P_r \left[ \frac{\partial H_r}{\partial Q_1} 5 \lambda_{ill}+
\frac{\partial H_r}{\partial Q_2} \left( 2Q_1 \lambda_{ill}+ 4 \lambda_{ia} \lambda_{all}
\right)+  \frac{\partial H_r}{\partial Q_3}  \left( 3Q_2 \lambda_{ill}+ 6 \lambda_{ia}^2
\lambda_{all}\right) \right] X_{r+1} \, , \\
0&=& \sum_{r=0}^3 2V_r^k \left[ \frac{\partial H_r}{\partial Q_1} 5 \lambda_{ill}+
\frac{\partial H_r}{\partial Q_2} \left( 2Q_1 \lambda_{ill}+ 4 \lambda_{ia} \lambda_{all}
\right)+  \frac{\partial H_r}{\partial Q_3} \left( 3Q_2 \lambda_{ill}+ 6 \lambda_{ia}^2
\lambda_{all}\right) \right] X_{r+1} \, ,
\end{eqnarray*}
with $Q_1= \lambda_{ll}$ , $Q_2= tr \lambda_{rs}^2$ , $Q_3= tr \lambda_{rs}^3$. \\
Now, for the first hypothesis at the beginning of this section, it follows that the
vectors  $\lambda_{ill}$  , $\lambda_{ia} \lambda_{all}$, $\lambda_{ia}^2 \lambda_{all}$
are linearly independent; consequently, the above relation becomes
\begin{eqnarray*}
0&=& \sum_{r=0}^3 P_r  \frac{\partial H_r}{\partial Q_s} X_{r+1} \, , \, 0= \sum_{r=0}^3
V_r^k  \frac{\partial H_r}{\partial Q_s}  \, , \, \mbox{for} \, s=1,2,3 .
\end{eqnarray*}
This result, for the second hypothesis at the beginning of this section, implies that
$\frac{\partial H_r}{\partial Q_s}=0$, that is, $H_r$ doesn't depend on $Q_1$, $Q_2$,
$Q_3$. Consequently it may depend only on $X_1$-$X_8$, as we desired to prove. \\
In this way, we have proved uniqueness only if the conditions 1) and 2), at the beginning
of this section, are satisfied. On the other hand, the set in which these conditions are
not satisfied is only a sub-manifold  of the domain; so our result on uniqueness must
hold in any case for continuity reasons. This can be clarified better with the following
example: If $F(x,y)$ is a continuous function such that
\begin{eqnarray*}
F(x,y)=
  \begin{cases}
    5 & \text{if} \, y \neq 0  \\
    f(x) & \text{if} \, y=0.
  \end{cases}
\end{eqnarray*}
then it follows $f(x)= F(x,0)= \lim_{y \rightarrow 0} F(x,y)=5$ so that $F(x,y)=5$ for
all values of $x,y$.
\section{The further condition (\ref{3}).}
We want now to impose the further condition $(\ref{3})$; we will see that it can be
nicely solved. The solution gives $H_0$, $H_1$, $H_2$, $H_3$, in terms of the arbitrary
functions $\psi=\psi ( X_1, \, X_2, \, X_3, \, X_4, \, X_5, \, Y_6, \, Y_7, \, Y_8)$,
$\varphi=\varphi (  X_1, \, X_2, \, X_3, \, X_4, \, Z_5, \, X_6, \, Z_7, \, Z_8)$,
$H_i^*=H_i^*(  X_1, \, X_2, \, X_3, \, X_4,\, Y_6, \, Y_7, \, Y_8)$ for $i$ going from  1
to 3, $H_j^{**}=H_j^{**}(  X_1, \, X_2, \, X_3, \, X_4, \, Z_5, \, Z_7, \, Z_8)$ for
$j=0,2,3$. This solution reads
\begin{eqnarray}\label{9}
  H_0 &=& \frac{1}{8} X_2 \left( \frac{\partial \psi}{\partial Y_6} + H_1^* \right) +
  \frac{1}{12} X_3 \left( \frac{\partial \psi}{\partial Y_7} + H_2^* \right) +
  \frac{1}{16} X_4 \left( \frac{\partial \psi}{\partial Y_8} + H_3^* \right) +
  \frac{\partial \psi}{\partial X_5} +   \\
  &{}& + X_2 \left( \frac{\partial \varphi}{\partial Z_5} + H_0^{**} \right) \, , \nonumber \\
H_1 &=& X_1 \left( \frac{\partial \psi}{\partial Y_6} + H_1^* \right) + \nonumber \\
  &{}& + 8 X_1 \left( \frac{\partial \varphi}{\partial Z_5} + H_0^{**} \right) -
  \frac{2}{3} X_3 \left( \frac{\partial \varphi}{\partial Z_7} + H_2^{**} \right) -
  \frac{1}{2} X_4 \left( \frac{\partial \varphi}{\partial Z_8} + H_3^{**} \right) +
   \frac{\partial \varphi}{\partial X_6}  \, ,  \nonumber \\
H_2 &=&  X_1 \left( \frac{\partial \psi}{\partial Y_7} + H_2^* \right)  + X_2 \left(
\frac{\partial \varphi}{\partial Z_7} + H_2^{**} \right) \, , \nonumber \\
H_3 &=& X_1 \left( \frac{\partial \psi}{\partial Y_8} + H_3^* \right) + X_2 \left(
\frac{\partial \varphi}{\partial Z_8} + H_3^{**} \right) \, , \nonumber
\end{eqnarray}
where it is understood that the right hand sides are calculated in
\begin{eqnarray}\label{10}
&{}&  Y_6= X_1X_6+ \frac{1}{8} \, X_2X_5 \quad , \quad Y_7= X_1X_7+ \frac{1}{12} \,
X_3X_5 \quad , \quad
  Y_8= X_1X_8+ \frac{1}{16} \, X_4X_5 \quad , \\
&{}&   Z_5= X_2X_5+ 8 \, X_1X_6 \quad , \quad Z_7= X_2X_7- \frac{2}{3} \, \, \, X_3X_6
\quad , \quad
  Z_8= X_2X_8- \frac{1}{2} \, X_4X_6 \quad . \nonumber
 \end{eqnarray}
In order to prove this result, let us start by noting that from  (\ref{6}) and (\ref{7})
it follows that $V^k_0$, $V^k_1$, $V^k_2$, $V^k_3$ don' t depend on $\lambda$ and,
moreover,
\begin{eqnarray}\label{11}
&{}& \frac{\partial X_1}{\partial \lambda_k} =0 \quad \quad  , \quad \frac{\partial
X_2}{\partial \lambda_k} =0 \quad \quad \, , \quad \frac{\partial X_3}{\partial
\lambda_k} =0 \quad \quad , \quad
\frac{\partial X_4}{\partial \lambda_k} =0 \quad , \\
&{}& \frac{\partial X_5}{\partial \lambda_k} =2 V^k_0 \, , \, \quad \frac{\partial
X_6}{\partial \lambda_k} = 2V^k_1 \quad , \quad
\frac{\partial X_7}{\partial \lambda_k} =2V^k_2 \quad , \quad   \frac{\partial X_8}{\partial \lambda_k} =2V^k_3 \quad , \nonumber \\
&{}& \frac{\partial X_1}{\partial \lambda} =0 \quad \quad , \quad \frac{\partial
X_2}{\partial \lambda} =0 \quad \quad \, , \quad \frac{\partial X_3}{\partial \lambda} =0
\quad \quad , \quad
\frac{\partial X_4}{\partial \lambda} =0 \quad , \nonumber \\
&{}& \frac{\partial X_5}{\partial \lambda} =16X_1 \, , \quad \frac{\partial X_6}{\partial
\lambda_k} = -2X_2 \, , \quad \frac{\partial X_7}{\partial \lambda} =- \frac{4}{3} X_3 \,
, \quad \frac{\partial X_8}{\partial \lambda} =-X_4 \quad . \nonumber
\end{eqnarray}
From $(\ref{11})_{9-12}$ we have also that the  coefficients of $H_0$, $H_1$, $H_2$,
$H_3$ in $h'$ don' t depend on $\lambda$; consequently, eq. $(\ref{3})$ becomes
\begin{eqnarray*}
 &{}& 8X_1 \left(  2 \frac{\partial H_0}{\partial X_5} V^k_0 +  2 \frac{\partial H_0}{\partial
X_6} V^k_1 +
2 \frac{\partial H_0}{\partial X_7} V^k_2 + 2 \frac{\partial H_0}{\partial X_8} V^k_3 \right) + \\
 &{}& -X_2 \left(  2 \frac{\partial H_1}{\partial X_5} V^k_0 +  2 \frac{\partial
H_1}{\partial X_6} V^k_1 +
2 \frac{\partial H_1}{\partial X_7} V^k_2 + 2 \frac{\partial H_1}{\partial X_8} V^k_3 \right) + \\
 &{}& -\frac{2}{3}X_3   \left(  2 \frac{\partial H_2}{\partial X_5} V^k_0 +  2 \frac{\partial
H_2}{\partial X_6} V^k_1 + 2 \frac{\partial H_2}{\partial X_7} V^k_2 + 2 \frac{\partial
H_2}{\partial X_8} V^k_3 \right) + \\
 &{}& - \frac{1}{2}X_4    \left(  2 \frac{\partial H_3}{\partial X_5} V^k_0 +  2 \frac{\partial
H_3}{\partial X_6} V^k_1 + 2 \frac{\partial H_3}{\partial X_7} V^k_2 + 2 \frac{\partial
H_3}{\partial X_8} V^k_3 \right) = \\
 &{}& = \frac{\partial H_0}{\partial \lambda} V^k_0 +   \frac{\partial H_1}{\partial
\lambda} V^k_1 +  \frac{\partial H_2}{\partial \lambda} V^k_2 +  \frac{\partial
H_3}{\partial \lambda} V^k_3 \, ,
\end{eqnarray*}
or,
\begin{eqnarray*}
&{}& \frac{\partial H_0}{\partial \lambda} = 16X_1   \frac{\partial H_0}{\partial X_5} -2
X_2 \frac{\partial H_1}{\partial X_5} - \frac{4}{3} X_3 \frac{\partial H_2}{\partial X_5}
- X_4 \frac{\partial H_3}{\partial X_5} \, , \\
&{}& \frac{\partial H_1}{\partial \lambda} = 16X_1   \frac{\partial H_0}{\partial X_6} -2
X_2 \frac{\partial H_1}{\partial X_6} - \frac{4}{3} X_3 \frac{\partial H_2}{\partial X_6}
- X_4 \frac{\partial H_3}{\partial X_6} \, ,  \\
&{}& \frac{\partial H_2}{\partial \lambda} = 16X_1   \frac{\partial H_0}{\partial X_7} -2
X_2 \frac{\partial H_1}{\partial X_7} - \frac{4}{3} X_3 \frac{\partial H_2}{\partial X_7}
- X_4 \frac{\partial H_3}{\partial X_7}  \, , \\
&{}& \frac{\partial H_3}{\partial \lambda} = 16X_1   \frac{\partial H_0}{\partial X_8} -2
X_2 \frac{\partial H_1}{\partial X_8} - \frac{4}{3} X_3 \frac{\partial H_2}{\partial X_8}
- X_4 \frac{\partial H_3}{\partial X_8} \, .
\end{eqnarray*}
These equations, for  $(\ref{11})_{13-16}$ become
\begin{eqnarray}\label{12}
&{}&  -2 X_2 \frac{\partial H_0}{\partial X_6} - \frac{4}{3} X_3 \frac{\partial
H_0}{\partial X_7} - X_4 \frac{\partial H_0}{\partial X_8} = -2 X_2 \frac{\partial
H_1}{\partial X_5} - \frac{4}{3} X_3 \frac{\partial H_2}{\partial X_5} - X_4
\frac{\partial H_3}{\partial X_5}  \, , \\
&{}& 16X_1   \frac{\partial H_1}{\partial X_5} - \frac{4}{3} X_3 \frac{\partial
H_1}{\partial X_7} - X_4 \frac{\partial H_1}{\partial X_8}= 16X_1   \frac{\partial
H_0}{\partial X_6} - \frac{4}{3} X_3 \frac{\partial H_2}{\partial X_6}
- X_4 \frac{\partial H_3}{\partial X_6}  \, , \nonumber \\
&{}& 16X_1   \frac{\partial H_2}{\partial X_5} -2 X_2 \frac{\partial H_2}{\partial X_6} -
 X_4 \frac{\partial H_2}{\partial
X_8}= 16X_1   \frac{\partial H_0}{\partial X_7} -2 X_2 \frac{\partial H_1}{\partial X_7}
- X_4 \frac{\partial H_3}{\partial X_7}   \, , \nonumber \\
&{}& 16X_1   \frac{\partial H_3}{\partial X_5} -2 X_2 \frac{\partial H_3}{\partial X_6} -
\frac{4}{3} X_3 \frac{\partial H_3}{\partial X_7} = 16X_1   \frac{\partial H_0}{\partial
X_8} -2 X_2 \frac{\partial H_1}{\partial X_8} - \frac{4}{3} X_3 \frac{\partial
H_2}{\partial X_8}  \, .  \nonumber
\end{eqnarray}
To find the solution of these equations, let us distinguish two cases:
\subsection{The case X$_1\neq 0$.}
From  $(\ref{12})_{2-4}$ we obtain
\begin{eqnarray}\label{13}
&{}& \frac{\partial H_0}{\partial X_6} = \frac{\partial H_1}{\partial X_5} - \frac{1}{12}
\frac{X_3}{X_1} \frac{\partial H_1}{\partial X_7} - \frac{1}{16} \frac{X_4}{X_1}
\frac{\partial H_1}{\partial X_8} + \frac{1}{12} \frac{X_3}{X_1} \frac{\partial
H_2}{\partial X_6} + \frac{1}{16} \frac{X_4}{X_1} \frac{\partial H_3}{\partial X_6} \, , \\
&{}& \frac{\partial H_0}{\partial X_7} = \frac{\partial H_2}{\partial X_5} - \frac{1}{8}
\frac{X_2}{X_1} \frac{\partial H_2}{\partial X_6} - \frac{1}{16} \frac{X_4}{X_1}
\frac{\partial H_2}{\partial X_8} + \frac{1}{8} \frac{X_2}{X_1} \frac{\partial
H_1}{\partial X_7} + \frac{1}{16} \frac{X_4}{X_1} \frac{\partial H_3}{\partial X_7} \, , \nonumber \\
&{}& \frac{\partial H_0}{\partial X_8} = \frac{\partial H_3}{\partial X_5} - \frac{1}{8}
\frac{X_2}{X_1} \frac{\partial H_3}{\partial X_6} - \frac{1}{12} \frac{X_3}{X_1}
\frac{\partial H_3}{\partial X_7}+ \frac{1}{8} \frac{X_2}{X_1} \frac{\partial
H_1}{\partial X_8} + \frac{1}{12} \frac{X_3}{X_1} \frac{\partial H_2}{\partial X_8} \, .
\nonumber
\end{eqnarray}
By substituting these expressions of the derivatives of $H_0$ in $(\ref{12})_{1}$, this
first equation becomes an identity. Let us now change functions and independent variables
according to the following relation
\begin{eqnarray}\label{14}
&{}& H_i = X_1 \tilde{H}_i ( X_1,  X_2,  X_3,  X_4,  X_5, \underbrace{X_1X_6+ \frac{1}{8}
\, X_2X_5} , \underbrace{ X_1X_7+ \frac{1}{12} \, X_3X_5 } , \underbrace{X_1X_8+
\frac{1}{16} \, X_4X_5} ) \, . \nonumber \\
&{}&  \quad \quad \quad \quad \quad \quad\quad \quad \quad \quad \quad \quad\quad \quad
\quad \quad \quad Y_6  \quad \quad \quad\quad \quad \quad \quad Y_7 \quad \quad
\quad\quad \quad \quad\quad Y_8
\end{eqnarray}
for $i=0, \cdots , 3$. With this change, eqs. (\ref{13}) become
\begin{eqnarray*}
&{}& \frac{\partial \tilde{H}_1}{\partial X_5} = \frac{\partial }{\partial Y_6} \left(
X_1 \tilde{H}_0 - \frac{1}{8} X_2 \tilde{H}_1 - \frac{1}{12} X_3 \tilde{H}_2 -
\frac{1}{16} X_4 \tilde{H}_3 \right) \, , \\
&{}& \frac{\partial \tilde{H}_2}{\partial X_5} = \frac{\partial }{\partial Y_7} \left(
X_1 \tilde{H}_0 - \frac{1}{8} X_2 \tilde{H}_1 - \frac{1}{12} X_3 \tilde{H}_2 -
\frac{1}{16} X_4 \tilde{H}_3 \right) \, , \\
&{}& \frac{\partial \tilde{H}_3}{\partial X_5} = \frac{\partial }{\partial Y_8} \left(
X_1 \tilde{H}_0 - \frac{1}{8} X_2 \tilde{H}_1 - \frac{1}{12} X_3 \tilde{H}_2 -
\frac{1}{16} X_4 \tilde{H}_3 \right) \, .
\end{eqnarray*}
So it will suffice to define  $\psi$ from
\begin{eqnarray*}
X_1 \tilde{H}_0 - \frac{1}{8} X_2 \tilde{H}_1 - \frac{1}{12} X_3 \tilde{H}_2 -
\frac{1}{16} X_4 \tilde{H}_3 = \frac{\partial \psi}{\partial X_5}
\end{eqnarray*}
to obtain, tanks to eqs. (\ref{14}), the result (\ref{9}), but with $\varphi=0$,
$H_j^{**}=0$. On the other hand, from  $(\ref{3})$ we see that the sum of two solutions
is still a solution. Consequently, it will suffice now to prove that (\ref{9}) is a
solution also with $\varphi \neq 0$, $H_j^{**} \neq 0$, $\psi =0$, $H^*_i=0$; this will
be the result of the following case.
\subsection{The case X$_2 \neq 0$.}
From eq.   $(\ref{12})_{1,3,4}$ we obtain
\begin{eqnarray}\label{15}
&{}& \frac{\partial H_1}{\partial X_5} = \frac{\partial H_0}{\partial X_6} + \frac{2}{3}
\frac{X_3}{X_2} \frac{\partial H_0}{\partial X_7} + \frac{1}{2} \frac{X_4}{X_2}
\frac{\partial H_0}{\partial X_8} - \frac{2}{3} \frac{X_3}{X_2} \frac{\partial
H_2}{\partial X_5} - \frac{1}{2} \frac{X_4}{X_2} \frac{\partial H_3}{\partial X_5} \, , \\
&{}& \frac{\partial H_1}{\partial X_7} = -8 \frac{X_1}{X_2} \frac{\partial H_2}{\partial
X_5} + \frac{\partial H_2}{\partial X_6} + \frac{1}{2} \frac{X_4}{X_2} \frac{\partial
H_2}{\partial X_8} + 8 \frac{X_1}{X_2} \frac{\partial
H_0}{\partial X_7} - \frac{1}{2} \frac{X_4}{X_2} \frac{\partial H_3}{\partial X_7} \, , \nonumber \\
&{}& \frac{\partial H_1}{\partial X_8} = -8 \frac{X_1}{X_2} \frac{\partial H_3}{\partial
X_5} + \frac{\partial H_3}{\partial X_6} + \frac{2}{3} \frac{X_3}{X_2} \frac{\partial
H_3}{\partial X_7} + 8 \frac{X_1}{X_2} \frac{\partial H_0}{\partial X_8} - \frac{2}{3}
 \frac{X_3}{X_2} \frac{\partial H_2}{\partial X_8} \, . \nonumber
\end{eqnarray}
By substituting these in $(\ref{12})_{2}$, this relation becomes an identity. Let us now
change functions and independent variables according to
\begin{eqnarray}\label{16}
&{}& H_i = X_2 \tilde{H}_i ( X_1,  X_2,  X_3,  X_4,  \underbrace{X_2X_5+ 8 \, X_1X_6  } ,
X_6 , \underbrace{X_2X_7- \frac{2}{3} \, \, \, X_3X_6 } , \underbrace{X_2X_8- \frac{1}{2} \, X_4X_6} ) \, . \nonumber \\
&{}&  \quad \quad \quad \quad \quad \quad\quad \quad \quad \quad \quad \quad\quad \quad
\quad  Z_5  \quad \quad \quad \quad \quad\quad \quad \quad \quad Z_7 \quad \quad
\quad\quad \quad \quad\quad Z_8
\end{eqnarray}
for $i=0, \cdots , 3$. With this change, eqs.  (\ref{15}) become
\begin{eqnarray*}
&{}& \frac{\partial \tilde{H}_0}{\partial X_6} = \frac{\partial }{\partial Z_5} \left( -8
X_1 \tilde{H}_0 + X_2 \tilde{H}_1 + \frac{2}{3} X_3 \tilde{H}_2 +
\frac{1}{2} X_4 \tilde{H}_3 \right) \, , \\
&{}& \frac{\partial \tilde{H}_2}{\partial X_6} = \frac{\partial }{\partial Z_7} \left(
 -8
X_1 \tilde{H}_0 + X_2 \tilde{H}_1 + \frac{2}{3} X_3 \tilde{H}_2 +
\frac{1}{2} X_4 \tilde{H}_3 \right) \, , \\
&{}& \frac{\partial \tilde{H}_3}{\partial X_6} = \frac{\partial }{\partial Z_8} \left(
 -8
X_1 \tilde{H}_0 + X_2 \tilde{H}_1 + \frac{2}{3} X_3 \tilde{H}_2 + \frac{1}{2} X_4
\tilde{H}_3 \right) \, .
\end{eqnarray*}
So it will suffice to define  $\varphi$ from
\begin{eqnarray*}
 -8
X_1 \tilde{H}_0 + X_2 \tilde{H}_1 + \frac{2}{3} X_3 \tilde{H}_2 + \frac{1}{2} X_4
\tilde{H}_3 = \frac{\partial \varphi}{\partial X_6}
\end{eqnarray*}
to obtain, thanks to eqs. (\ref{16}), the eqs. (\ref{9}), but with $\psi =0$, $H^*_i=0$,
as afore said.
\section{The convexity of $h'$.}
In order that our system $(\ref{1})$ be hyperbolic, we have now to impose that the
hessian matrix  $\frac{\partial^2 h'}{\partial \lambda_A
\partial \lambda_B}$ is positive defined, with  $\lambda_A$ the generic component of the Lagrange
multipliers. In other words, the quadratic form $Q=\frac{\partial^2 h'}{\partial
\lambda_A
\partial \lambda_B} \delta \lambda_A \delta \lambda_B$ has to be positive definite. Let
us exploit this with the potentials $(\ref{8})_1$; in these expressions, except for
replacing  $X_i$ with $X_i/(X_1)$ for $i=5, \cdots , 8$, the remaining polynomials in
$X_j/(X_1)$ for $j=2, \cdots , 4$ have been chosen in order to eliminate from $X_i$ for
$i=5, \cdots , 8$ the terms depending only on $\lambda_{ab}$. \\
Well, we want now to evaluate this quadratic form  $Q$ in the state, which will be called
$C$, where $\lambda_i=0$, $\lambda_{ij}= \frac{1}{3} \lambda_{ll} \delta_{ij}$,
$\lambda_{ill}=0$; so there remain, as independent variables $\lambda$, $\lambda_{ll}$,
$\lambda_{ppll}$. This is an intermediate state with respect to equilibrium, where we
have also $\lambda_{ppll}=0$. To this end we need the expressions of our variables up to
second order with respect to the state $C$. After some calculations, we find
\begin{eqnarray*}
&{}& \eta_1 = X_1 = \lambda_{ppll} \, , \\
&{}& \eta_2 =  \frac{2}{\lambda_{ppll}}  \lambda_{all}
\lambda_{all}-  \frac{16}{5}\lambda_{ll} \, ,    \\
&{}& \eta_3 \simeq \frac{32}{75} \lambda_{ll}^2 - 4 (tr \lambda^{2}_{<ab>}) -
\frac{8}{15}
\frac{\lambda_{ll}}{\lambda_{ppll}} \lambda_{all} \lambda_{all} \, , \\
&{}& \eta_4 \simeq - \frac{64}{27 \cdot 125} \lambda_{ll}^3 + \frac{8}{15} \lambda_{ll}
(tr \lambda^{2}_{<ab>}) + \frac{8}{25 \cdot 9} \frac{\lambda_{ll}^2}{\lambda_{ppll}}
\lambda_{all} \lambda_{all} \, , \\
&{}& \eta_5 \simeq 16 \lambda - 4 \lambda_a \frac{\lambda_{all}}{\lambda_{ppll}} +
\frac{1}{3} \frac{\lambda_{ll}}{\lambda_{ppll}^2} \lambda_{all} \lambda_{all} \, , \\
&{}& \eta_6 \simeq - \frac{32}{5} \lambda \lambda_{ll} + \left(
\frac{4\lambda}{\lambda_{ppll}} - \frac{1}{45} \frac{\lambda_{ll}^2}{\lambda_{ppll}^2}
\right) \lambda_{all} \lambda_{all} + 4 \lambda_{a} \lambda_{a} + \frac{4}{15}
\frac{\lambda_{ll}}{\lambda_{ppll}} \lambda_{a} \lambda_{all} \, , \\
&{}& \eta_7 \simeq \frac{64}{75} \lambda \lambda_{ll}^2 + \frac{56}{15}
\lambda_{ll}\lambda_{a}\lambda_{a} + \frac{32}{9 \cdot 25}
\frac{\lambda_{ll}^2}{\lambda_{ppll}} \lambda_{a} \lambda_{all} - 8 \lambda (tr
\lambda^{2}_{<ab>}) + \\
&{}& \quad \quad \quad - \frac{12}{5} \frac{\lambda}{\lambda_{ppll}} \lambda_{ll}
\lambda_{all} \lambda_{all} - \frac{47}{27 \cdot 125}
\frac{\lambda_{ll}^3}{\lambda_{ppll}^2} \lambda_{all} \lambda_{all} \, , \\
&{}& \eta_8 \simeq - \frac{8 \cdot 16}{27 \cdot 125} \lambda \lambda_{ll}^3 + \frac{16}{9
\cdot 25} \lambda_{ll}^2 \lambda_{a}\lambda_{a} - \frac{16}{9 \cdot 125}
\frac{\lambda_{ll}^3}{\lambda_{ppll}} \lambda_{a} \lambda_{all} + \frac{16}{15} \lambda
\lambda_{ll}  (tr
\lambda^{2}_{<ab>}) + \\
&{}& \quad \quad \quad + \frac{16}{9 \cdot 25} \frac{\lambda
\lambda_{ll}^2}{\lambda_{ppll}}  \lambda_{all} \lambda_{all} +  \frac{4}{9 \cdot 625}
\frac{\lambda_{ll}^3}{\lambda_{ppll}^2} \lambda_{all} \lambda_{all} \, ,
\end{eqnarray*}
with $\lambda_{<ab>}= \lambda_{ab}- \frac{1}{3} \lambda_{ll} \delta_{ij}$. \\
After that, from $h'=h'(\eta_i)$, we find that the expression of $h'$ up to second order
with respect to the state  $C$ is
\begin{eqnarray*}
&{}& h' \simeq h' ( \eta_i^*) + \sum_{j=2}^8 \left( \frac{\partial h'}{\partial \eta_j}
\right)^* ( \eta_j - \eta_j^*) = \\
&{}& = h'  ( \eta_i^*) + \left( \frac{\partial h'}{\partial \eta_2} \right)^*  \left(
\frac{2}{\lambda_{ppll}}  \lambda_{all}
\lambda_{all} \right) +    \\
&{}& + \left( \frac{\partial h'}{\partial \eta_3} \right)^*  \left(  - 4 (tr
\lambda^{2}_{<ab>}) - \frac{8}{15}
\frac{\lambda_{ll}}{\lambda_{ppll}} \lambda_{all} \lambda_{all} \right) + \\
&{}& + \left( \frac{\partial h'}{\partial \eta_4} \right)^*  \left(  \frac{8}{15}
\lambda_{ll} (tr \lambda^{2}_{<ab>}) + \frac{8}{25 \cdot 9}
\frac{\lambda_{ll}^2}{\lambda_{ppll}}
\lambda_{all} \lambda_{all} \right) + \\
&{}& + \left( \frac{\partial h'}{\partial \eta_5} \right)^*  \left(  - 4 \lambda_a
\frac{\lambda_{all}}{\lambda_{ppll}} +
\frac{1}{3} \frac{\lambda_{ll}}{\lambda_{ppll}^2} \lambda_{all} \lambda_{all} \right) + \\
&{}& + \left( \frac{\partial h'}{\partial \eta_6} \right)^* \left[
 \left( \frac{4\lambda}{\lambda_{ppll}} - \frac{1}{45}
\frac{\lambda_{ll}^2}{\lambda_{ppll}^2} \right) \lambda_{all} \lambda_{all} + 4
\lambda_{a} \lambda_{a} + \frac{4}{15}
\frac{\lambda_{ll}}{\lambda_{ppll}} \lambda_{a} \lambda_{all} \right] + \\
&{}& + \left( \frac{\partial h'}{\partial \eta_7} \right)^*  \left(
 \frac{56}{15} \lambda_{ll}\lambda_{a}\lambda_{a} + \frac{32}{9 \cdot 25}
\frac{\lambda_{ll}^2}{\lambda_{ppll}} \lambda_{a} \lambda_{all} - 8 \lambda (tr
\lambda^{2}_{<ab>}) + \right.\\
&{}&  \left. \quad \quad \quad - \frac{12}{5} \frac{\lambda}{\lambda_{ppll}} \lambda_{ll}
\lambda_{all} \lambda_{all} - \frac{47}{27 \cdot 125}
\frac{\lambda_{ll}^3}{\lambda_{ppll}^2} \lambda_{all} \lambda_{all} \right) + \\
&{}& + \left( \frac{\partial h'}{\partial \eta_8} \right)^*  \left( \frac{16}{9 \cdot 25}
\lambda_{ll}^2 \lambda_{a}\lambda_{a} - \frac{16}{9 \cdot 125}
\frac{\lambda_{ll}^3}{\lambda_{ppll}} \lambda_{a} \lambda_{all} + \frac{16}{15} \lambda
\lambda_{ll}  (tr
\lambda^{2}_{<ab>}) + \right. \\
&{}& \quad \quad \quad \left. + \frac{16}{9 \cdot 25} \frac{\lambda
\lambda_{ll}^2}{\lambda_{ppll}}  \lambda_{all} \lambda_{all} +  \frac{4}{9 \cdot 625}
\frac{\lambda_{ll}^3}{\lambda_{ppll}^2} \lambda_{all} \lambda_{all} \right) \, ,
\end{eqnarray*}
where the apex * denotes a quantity calculated in the state  $C$, so that we have also
\begin{eqnarray*}
&{}& \eta_1^* = X_1 = \lambda_{ppll} \quad , \quad
 \eta_2^* =  -  \frac{16}{5}\lambda_{ll} \quad  , \quad
\eta_3^* = \frac{32}{75} \lambda_{ll}^2  \quad  , \quad
 \eta_4^* =  - \frac{64}{27 \cdot 125} \lambda_{ll}^3  \, , \\
&{}& \eta_5^* = 16 \lambda  \quad  , \quad
 \eta_6^*=  - \frac{32}{5} \lambda \lambda_{ll} \quad  , \quad
\eta_7^* = \frac{64}{75} \lambda \lambda_{ll}^2  \quad  , \quad  \eta_8^* = - \frac{8
\cdot 16}{27 \cdot 125} \lambda \lambda_{ll}^3  \, .
\end{eqnarray*}
Taking into account these intermediate results and the additivity of $Q=\frac{\partial^2
h'}{\partial \lambda_A \partial \lambda_B} \delta \lambda_A \delta \lambda_B$ and by
calculating $\frac{\partial^2 h'}{\partial \lambda_A
\partial \lambda_B} $ in the confront state  $C$, we find
\begin{eqnarray*}
  Q= Q_1 + Q_2 + Q_3 \, ,  \quad \mbox{with}
\end{eqnarray*}
\begin{eqnarray*}
Q_1 = a_{11} ( \delta \lambda)^2+  2a_{12} \delta \lambda \delta \lambda_{ll} +
2a_{13}\delta \lambda \delta \lambda_{ppll}+ a_{22} (\delta \lambda_{ll})^2+ 2a_{23}
\delta\lambda_{ll}\delta\lambda_{ppll}+ a_{33}(\delta \lambda_{ppll})^2 \, ,
\end{eqnarray*}
\begin{eqnarray*}
Q_2 = b_{11} \left( \frac{\delta \lambda_{all}}{ \lambda_{ppll}} \right) \cdot \left(
\frac{\delta \lambda_{all}}{ \lambda_{ppll}} \right) + 2b_{12} \left( \frac{\delta
\lambda_{all}}{ \lambda_{ppll}} \right) \delta \lambda_{a} + b_{22} \delta \lambda_{a}
\delta \lambda_{a} \, ,
\end{eqnarray*}
\begin{eqnarray*}
Q_3 =   c  \, \, \delta \lambda_{<rs>} \delta \lambda_{<rs>} \, ,
\end{eqnarray*}
\begin{eqnarray*}
&{}& a_{11} = \frac{\partial^2 h' (\eta_i^*)}{\partial \lambda^2} \, , \, a_{12} =
\frac{\partial^2 h' (\eta_i^*)}{\partial \lambda \partial \lambda_{ll} } \, , \, a_{13} =
\frac{\partial^2 h' (\eta_i^*)}{\partial \lambda \partial \lambda_{ppll} } \, , \\
&{}& a_{22} = \frac{\partial^2 h' (\eta_i^*)}{\partial \lambda_{ll}^2} \, , \, a_{23} =
\frac{\partial^2 h' (\eta_i^*)}{\partial \lambda_{ll} \partial \lambda_{ppll} } \, , \,
a_{33} = \frac{\partial^2 h' (\eta_i^*)}{\partial  \lambda_{ppll}^2 } \, ,
\end{eqnarray*}
\begin{eqnarray*}
&{}& b_{11} = \frac{2}{3} \lambda_{ll} \left( \frac{\partial h'}{\partial \eta_5}
\right)^* - \frac{2}{45} \lambda_{ll}^2 \left( \frac{\partial h'}{\partial \eta_6}
\right)^* - 2 \frac{47}{27 \cdot 125} \lambda_{ll}^3 \left( \frac{\partial h'}{\partial
\eta_7} \right)^* + \frac{8}{9 \cdot 625} \lambda_{ll}^3 \left( \frac{\partial h'}{\partial \eta_8} \right)^*  + \\
&{}& \quad \quad + 2 \lambda_{ppll} \left[ 2 \left( \frac{\partial h'}{\partial \eta_2}
\right)^* - \frac{8}{15} \lambda_{ll} \left( \frac{\partial h'}{\partial \eta_3}
\right)^* + \frac{8}{25 \cdot 9}\lambda_{ll}^2 \left( \frac{\partial h'}{\partial \eta_4}
\right)^* + 4 \lambda \left( \frac{\partial h'}{\partial \eta_6} \right)^* + \right. \\
&{}& \left. \quad \quad - \frac{12}{5} \lambda \lambda_{ll} \left( \frac{\partial
h'}{\partial \eta_7} \right)^* + \frac{16}{9 \cdot 25} \lambda \lambda_{ll}^2 \left(
\frac{\partial h'}{\partial \eta_8} \right)^* \right] \, , \\
&{}& b_{22} = 8 \left( \frac{\partial h'}{\partial \eta_6} \right)^* + \frac{112}{15}
\lambda_{ll} \left( \frac{\partial h'}{\partial \eta_7} \right)^* + \frac{32}{9 \cdot 25}
\lambda_{ll}^2 \left( \frac{\partial h'}{\partial \eta_8} \right)^*  \, , \\
&{}& b_{12} = -4 \left( \frac{\partial h'}{\partial \eta_5} \right)^* + \frac{4}{15}
\lambda_{ll} \left( \frac{\partial h'}{\partial \eta_6} \right)^* + \frac{32}{9 \cdot 25}
\lambda_{ll}^2 \left( \frac{\partial h'}{\partial \eta_7} \right)^* - \frac{16}{9 \cdot
125} \lambda_{ll}^3 \left( \frac{\partial h'}{\partial \eta_8} \right)^* \, , \\
&{}& c = -8 \left( \frac{\partial h'}{\partial \eta_3} \right)^* + \frac{16}{15}
\lambda_{ll} \left( \frac{\partial h'}{\partial \eta_4} \right)^* - 16 \lambda \left(
\frac{\partial h'}{\partial \eta_7} \right)^* + \frac{32}{15} \lambda \lambda_{ll} \left(
\frac{\partial h'}{\partial \eta_8} \right)^* \, .
\end{eqnarray*}
Consequently, the required convexity holds if
\begin{eqnarray*}
&{}& a_{11}>0 \quad , \quad \left| \begin{matrix}
  a_{11} & a_{12} \\
  a_{21} & a_{22}
\end{matrix} \right| >0 \quad , \quad \left| \begin{matrix}
  a_{11} & a_{12} & a_{13} \\
  a_{21} & a_{22} & a_{23} \\
  a_{31} & a_{32} & a_{33}
\end{matrix} \right| >0 \quad , \\
&{}& b_{11}>0 \quad , \quad \left| \begin{matrix}
  b_{11} & b_{12} \\
  b_{21} & b_{22}
\end{matrix} \right| >0 \quad , \quad c >0 \quad .
\end{eqnarray*}
We note that these conditions are continuous in $\lambda_{ppll}$, so that we may impose
them only calculated  in $\lambda_{ppll}=0$; in this way we will obtain the requested
convexity not only in a neighborhood of the state  $C$, but also in a neighborhood of
equilibrium. \\
We have performed the same passages also by starting from eqs. $(\ref{5})_1$, instead of
$(\ref{8})_1$; in this way we have found that  $Q$ isn't positive defined. We conclude
that only eq.  $(\ref{8})$ is the correct expression to use. \\
We note that also the
results of the previous section can be written taking into account the expression
$(\ref{8})_1$. In particular, we can use the expressions at the end of section 1 to find
$X_1$-$X_8$ as functions of $\eta_1$-$\eta_8$. After that, by using also eqs.
$(\ref{10})$, we can obtain $Y_5$-$Y_8$, with $Y_5=X_5$, that is
\begin{eqnarray}\label{ange1}
&{}&  Y_5= \eta_1 \eta_5 - \frac{1}{2} \eta_3 + \frac{3}{64} (\eta_2)^2 \, , \\
&{}&  Y_6 = \eta_1  \left[ \eta_1 \eta_6 - \frac{1}{2} \eta_4+ \frac{1}{8} \eta_1 \eta_2
\eta_5 + \frac{1}{256} (\eta_2)^2\right] \, , \nonumber \\
&{}&  Y_7 = \eta_1  \left[ \eta_1 \eta_7 + \frac{1}{16} \eta_2 \eta_4+ \frac{1}{2^9}
\eta_3 (\eta_2)^2+ \frac{1}{12} \eta_1 \eta_3 \eta_5 - \frac{1}{24}  (\eta_3)^2\right] \, ,  \nonumber \\
&{}&  Y_8 = \eta_1  \left[ \eta_1 \eta_8 + \frac{1}{16} \eta_1 \eta_4 \eta_5+
\frac{1}{2^{10}} \eta_4 (\eta_2)^2- \frac{1}{32} \eta_3 \eta_4 \right] \, .  \nonumber
\end{eqnarray}
From $K_i=\eta_1 H_i$ and by defining $\vartheta= \eta_1 \psi$, $K_i^*=\frac{1}{\eta_1}
H_i^*$ for $i=1,2,3$, we can rewrite eqs. $(\ref{9})$. We will limit ourselves to the
case $X_1 \neq 0$, so that we have $\varphi=0$, $H_j^{**}=0$. The result is that the
solution gives $K_0$, $K_1$, $K_2$, $K_3$, in terms of the arbitrary functions
$\vartheta=\vartheta ( \eta_1, \, \eta_2, \, \eta_3, \, \eta_4, \, Y_5, \, Y_6, \, Y_7,
\, Y_8)$,  $K_i^*=K_i^*(  \eta_1, \, \eta_2, \, \eta_3, \, \eta_4,\, Y_6, \, Y_7, \,
Y_8)$ for $i$ going from  1 to 3. This solution reads
\begin{eqnarray}\label{ange2}
  K_0 &=& \frac{1}{8} \eta_1 \eta_2 \left( \frac{\partial \vartheta}{\partial Y_6} + K_1^* \right) +
  \frac{1}{12} \eta_1 \eta_3 \left( \frac{\partial \vartheta}{\partial Y_7} + K_2^* \right) +
  \frac{1}{16} \eta_1 \eta_4 \left( \frac{\partial \vartheta}{\partial Y_8} + K_3^* \right) +
  \frac{\partial \vartheta}{\partial Y_5}  \, , \nonumber \\
K_1 &=& \eta_1 \left( \frac{\partial \vartheta}{\partial Y_6} + K_1^* \right)   \, ,  \nonumber \\
K_2 &=&  \eta_1 \left( \frac{\partial \vartheta}{\partial Y_7} + K_2^* \right)  \, , \nonumber \\
K_3 &=& \eta_1 \left( \frac{\partial \vartheta}{\partial Y_8} + K_3^* \right)  \, ,
\nonumber
\end{eqnarray}
where it is understood that the right hand sides are calculated in (\ref{ange1}).
\section{The subsystems}
Other interesting particulars of our solution become manifest when we search the
subsystems of   $(\ref{1})$. \\
As example, eqs. $(\ref{4})$ calculated in $\lambda_{ill}=0$, $\lambda_{ppll}=0$ become
the conditions we would have by starting only with  $(\ref{1})_{1-3}$. But eq.
$(\ref{5})_2$ in $\lambda_{ill}=0$, $\lambda_{ppll}=0$ gives $h'=0$ which cannot be
accepted for the required convexity. This problem isn't avoided neither by using eqs.
(\ref{8}) because the consequent solutions don't satisfy the conditions  (\ref{4})
calculated for the subsystem. To verify that this is the case, it suffices to note that
$(\ref{4})_1$ with $\eta_5$ instead of  $h'$ is satisfied, but if we replace $\eta_5$
with its value in $\lambda_{ill}=0$, $\lambda_{ppll}=0$, that is $16 \lambda$, we see
that this satisfies no more eq. $(\ref{4})_1$ calculated in $\lambda_{ill}=0$,
$\lambda_{ppll}=0$! The reason is  evident from the fact that  $\eta_5$ satisfies eq.
$(\ref{4})_1$; but, if we calculate this equation in $\lambda_{ill}=0$, we find
\begin{eqnarray*}
0&=& \frac{\partial \eta_5}{\partial\lambda} \lambda_{i}+2\lambda_{ij}\frac{\partial
\eta_5}{\partial\lambda_{j}}+  4\lambda_{ppqq} \left( \frac{\partial
\eta_5}{\partial\lambda_{ill}} \right)_{\lambda_{ill}=0} \, , \quad \mbox{or}
\end{eqnarray*}
\begin{eqnarray*}
0&=& \frac{\partial \eta_5}{\partial\lambda} \lambda_{i}+2\lambda_{ij}\frac{\partial
\eta_5}{\partial\lambda_{j}}-16 \lambda_{i} \, ,
\end{eqnarray*}
whose value  in $\lambda_{ppll}=0$ isn't a solution of eq. $(\ref{4})_1$ calculated in
$\lambda_{ill}=0$, $\lambda_{ppll}=0$. \\
An  idea may be that to redo the passages of section 5 of paper
(\cite{3}) but starting from the beginning with $\lambda_{ill}=0$,
$\lambda_{ppll}=0$, that is, with
\begin{eqnarray*}
{\lambda^\beta}_\gamma&=& \frac{1}{m_0^2} \left[
 \begin{pmatrix}
  \frac{2}{3} \lambda_{ll} & 0_{j}  \\
    {} & {} \\
  0_{i} & \lambda_{ij} + \frac{1}{3} \lambda_{ll} \delta_{ij}
\end{pmatrix} +  \frac{1}{c^2}  \begin{pmatrix}
  - \lambda & 0_{j} \\
    {} & {} \\
  0_{i} & - \lambda \delta_{ij}
\end{pmatrix} \right]  \, ,  \\
&{}&  \\
{\lambda^\beta}&=& \frac{c}{m_0} \left[
  \begin{pmatrix}
  - \frac{2}{3} \lambda_{ll}  \\
    {}  \\
  0_{i}
\end{pmatrix} + \frac{1}{c}  \begin{pmatrix}
 0 \\
    {}\\
  \lambda_{i}
\end{pmatrix} \right]  \, ;
\end{eqnarray*}
but in this case, among the scalars there is the one coming from
$\lambda_\beta \lambda^\beta + \frac{1}{16} Q_1^2 m_0^2c^2$, that
is $\lambda_a \lambda_a - \frac{4}{3} \lambda \lambda_{ll}$ and
this, substituted to  $h'$ in  $(\ref{4})_1$ calculated in
$\lambda_{ill}=0$, $\lambda_{ppll}=0$, doesn't satisfy it. The
same thing can be said if we start from eq. (89) instead of (80),
both of  (\cite{3}). While, if we start from eq. (88) instead of
(80) (both of  (\cite{3})), we obtain quickly $h'=0$, $\phi'^k=0$.
In other words, the subsystem with 10 moments cannot be obtained
in any way as a non
relativistic limit. \\
What about the subsystem with 5 moments? \\
If in eq.(9) of  (\cite{3}) we substitute  $\lambda_{ij}=
\frac{1}{3} \lambda_{ll} \delta_{ij}$, $\lambda_{ill}=0$,
$\lambda_{ppll}=0$, they become the entropy principle for the
system constituted only by  $(\ref{1})_{1,2}$ and by the trace of
$(\ref{1})_{3}$, with Lagrange multipliers $\lambda$, $\lambda_i$,
$\frac{1}{3}
\lambda_{ll}$ respectively. \\
The equations  (\ref{4}) then become
\begin{eqnarray}\label{17}
0&=& \frac{\partial h'}{\partial\lambda} \lambda_{i}+\frac{2}{3}
\lambda_{ll}\frac{\partial h'}{\partial\lambda_{i}} \, , \, 0= \frac{\partial
\phi'^k}{\partial\lambda} \lambda_{i}+\frac{2}{3} \lambda_{ll}\frac{\partial
\phi'^k}{\partial\lambda_{i}} + h' \delta_{ik} \, .
\end{eqnarray}
With arguments like those above described, the solution of this equation cannot be found
from (\ref{5}), nor from (\ref{8}) calculated in the above values of $\lambda_{ij}$,
$\lambda_{ill}$, $\lambda_{ppll}$. \\
Instead of this, the idea of redoing the passages of section 5 of
(\cite{3}), but starting from the beginning with $\lambda_{ij}=
\frac{1}{3} \lambda_{ll} \delta_{ij}$, $\lambda_{ill}=0$,
$\lambda_{ppll}=0$, is successful . In fact, starting from (80) or
from (89) (both of (\cite{3})), we find
\begin{eqnarray*}
  h' = - \frac{2}{3} \lambda_{ll} H_0 \quad , \quad  \phi'^k= H_0 \lambda^k
\end{eqnarray*}
where $H_0$ is a function of  $\lambda_{ll}$, $\lambda_a \lambda_a - \frac{4}{3} \lambda
\lambda_{ll}$. These functions satisfy effectively eqs.  (\ref{17}). More than that, we
have that they satisfy automatically also eq. (\ref{3})! \\
Obviously, we cannot obtain this result by starting from the
beginning from (88) of (\cite{3}) because in this case we would
obtain quickly  $h'=0$, $\phi'^k=0$. On the other hand, if we
start from (88) of (\cite{3}) we obtain $\lambda^\beta=0$ and this
isn't adequate to describe the relativistic model; the less for its limit! \\
In other words, we have found that the following diagram isn't commutative \\
\begin{tabular}{ccccc}
  (relativistic model  & $\rightarrow$ & $\rightarrow$ & $\rightarrow$ & (relativistic subsystem with  \\
   with 14 moments) & ${}$ & ${}$ & ${}$ &  with 5 moments) \\
      ${}$ & ${}$ & ${}$ & ${}$ &  ${}$ \\
  $\downarrow$ non relativistic limit & {} & {} & {} & $\downarrow$ non relativistic limit \\
        ${}$ & ${}$ & ${}$ & ${}$ &  ${}$ \\
  (classical model  & ${}$ & (classical subsystem & ${}$ & (classical model
  \\
   with 14 moments) & $\rightarrow$ & with 5 moments) & $\neq$ &  with 5
  moments)\\
\end{tabular}
\\
${}$\\
${}$\\
 and this is quite different from the results obtained with expansions around
equilibrium,
that is, \\
${}$\\
${}$ \\
\begin{tabular}{ccccc}
  (classical model  & ${}$ & (classical subsystem & ${}$ & (classical model
  \\
   with 14 moments) & $\rightarrow$ & with 5 moments) & = &  with 5
  moments)\\
\end{tabular}
\\
${}$\\
${}$\\
even if this has been until now proved only for the less
restrictive case of ideal gases \cite{12} .

\end{document}